# Towards wafer scale fabrication of graphene based spin valve devices

*Ahmet Avsar[1,*], Tsung-Yeh Yang[2,3,*], Su-Kang Bae[4,*], Jayakumar Balakrishnan[1], Frank Volmer[2,3], Manu Jaiswal[1], Zheng Yi[1,4], Syed Rizwan Ali[2,3], Gernot Güntherodt[2,3], Byung-Hee Hong[4,†], Bernd Beschoten[2,3,†], and Barbaros Özyilmaz[1,5,6,†]*

[1]Graphene Research Center & Department of Physics, 2 Science Drive 3, National University of Singapore, Singapore 17542, Singapore

[2]II. Institute of Physics, RWTH Aachen University, 52074 Aachen, Germany

[3]JARA: Fundamentals of Future Information Technology, 52074 Aachen, Germany

[4]SKKU Advanced Institute of Nanotechnology (SAINT) and Center for Human Interface Nanotechnology (HINT), Sungkyunkwan University. Suwon, 440-746, Korea

[5]Nanocore, 4 Engineering Drive 3, National University of Singapore, Singapore 117576, Singapore

[6]NUS Graduate School for Integrative Sciences and Engineering (NGS), Singapore.

*These authors contributed equally to this work

[†]Corresponding authors: barbaros@nus.edu.sg, bernd.beschoten@physik.rwth-aachen.de, byunghee@skku.edu

**ABSTRACT** We demonstrate injection, transport and detection of spins in spin valve arrays patterned in both copper based chemical vapor deposition (Cu-CVD) synthesized wafer scale single layer (SLG) and bilayer graphene (BLG). We observe spin relaxation times comparable to those reported for exfoliated graphene samples demonstrating that CVD specific structural differences such as nano-ripples and grain boundaries do not limit spin transport in the present samples. Our observations make Cu-CVD graphene a promising material of choice for large scale spintronic applications.

High charge mobility,[1] small spin-orbit coupling,[2] negligible hyperfine interaction,[3] the electric field effect[4] and last but not least the ability to sustain large current densities[5] make graphene an exceptional material for spintronic applications. The demonstration of micrometer long spin relaxation length in exfoliated SLG and BLG even at room temperature (RT)[6]-[12] and spin relaxation times in the order of nanoseconds[11]-[12] may pave the way to realize several of the recently proposed spin based device concepts.[13]-[15] However, for realistic device applications it remains to be seen, if such impressive spin transport properties can also be achieved in wafer scale CVD graphene. Equally important, spin transport studies based on micromechanically exfoliated graphene sheets are often too slow for the quick exploration of the basic spin properties of graphene and for testing potential device architectures. The recent progress in the Cu-based CVD growth of graphene has a strong impact on charge based graphene device applications.[16] However, CVD graphene has a large number of structural differences when compared to exfoliated graphene such as grain boundaries,[17] defects like pentagons, heptagons, octagons, vacancies, 1D line charges[18] and in the case of bilayer graphene possibly interlayer stacking faults.[19]-[20] In addition, the current growth and transfer process introduces residual catalysts, wrinkles, quasi-periodic nanoripple arrays and new classes of organic residues.[19] Despite all of these defects, charge mobilities in CVD graphene field effect transistors (FETs) have been comparable to what has been reported for most exfoliated graphene FETs on Si/SiO$_2$ substrates.[21]

Whether this synthesis route will also play an important role for spin transport studies and large scale spin-based device applications depends on how the same defects affect the spin relaxation times.

In this Letter, we demonstrate spin transport in Cu-CVD grown SLG and BLG transferred onto conventional Si/SiO$_2$ substrates and discuss the role of nano-ripples, a ubiquitous surface structure of Cu-CVD graphene[19]. The growth and transfer of large-scale Cu-CVD graphene are the same as in Ref. (*17*). By controlling the post-growth annealing time of CVD graphene, we can obtain films with SLG coverage up to 95% or additional BLG coverage up to 40%. The latter samples are ideal for directly comparing spin transport in both systems. The inset in Fig. 1-a shows the optical image of CVD SLG and BLG on a Si/SiO$_2$ substrate. Raman spectra (Fig. 1-a) with insignificant D-band peak near 1400 cm$^{-1}$ show the high quality of both single-layer and A-B stacked bilayer samples. Spin valves are fabricated by first forming isolated SLG and BLG stripes by means of a PMMA etch mask. A second e-beam lithography step is used to form the device electrodes. Next, we deposit in the same run a ~ 2 nm thick MgO layer followed by 35 nm thick Co electrodes; details are discussed elsewhere.[11] This approach allows the batch-fabrication of large arrays of lateral spin-valves with a fast turn-around time well suited for studying device physics. An optical image of a 5 × 5 array of such devices is shown in Fig. 1-c (lower panel) together with a scanning electron microscopy (SEM) image (Fig. 1-c (upper panel)) of multiple spin-valve junctions showing the specific electrode configuration at a single site. The typical length and width of the spin channel in our spin valve devices are in the range of 1 µm to 2 µm. Measurements are performed with standard a.c. lock-in techniques at low frequencies using the local four terminal set-up for charge conductivity measurements and the non-local set-up for spin transport measurements.[22] The schematic of the non-local set-up is shown in Figure 1-b. The spin transport results obtained from CVD graphene are compared with the results from exfoliated graphene samples of similar charge mobilities prepared under identical conditions (see Supplementary Information).

Prior to any spin transport measurements, we characterized the conductivity of our devices as a function of back gate voltage at RT and at 5K. Figures 2-a and 2-f show the typical ambipolar field effect in our CVD SLG and BLG devices, respectively. A weak electron doping, possibly resulting from the MgO barrier, is observed in all our devices (carrier density $n \approx 0$ at $V_{\text{Dirac Point}} \approx -5$ V, not shown). In addition, our spin valves show a strong asymmetry between the electron ($n > 0$) and hole ($n < 0$) doped region, such that the conductance in the hole region is strongly distorted.[23] Here, we limit our spin transport analysis mainly to the electron doped region. In total, we have measured spin transport across 15 CVD SLG and BLG non-local spin-valve junctions. Field effect mobilities $\mu = \Delta\sigma/(e\Delta n)$ are extracted at $n \approx 2 \times 10^{12}/\text{cm}^2$, and vary from 1000 to 2100 cm$^2$/Vs. Here we discuss representative CVD SLG junctions and BLG junctions with mobilities of $\approx$ 1400 cm$^2$/Vs and $\approx$ 2100 cm$^2$/Vs, respectively. These values are similar to the values reported for most of the exfoliated graphene based spin valves in the literature.[6]-[12] Therefore, this allows a direct comparison of the spin transport properties of CVD graphene with exfoliated graphene.

We first discuss RT spin transport results in CVD SLG near the charge neutrality point (CNP). Sweeping the in-plane magnetic field $B_{\parallel}$ (Fig. 1-b) changes the relative magnetization directions of the Co electrodes and hence the spin accumulation between the injector and detector electrodes. This leads to a clear bi-polar non-local spin signal with a change in resistance of $\Delta R \approx 4$ Ω (Fig. 2-b). The origin of the spin signal is confirmed by conventional Hanle spin precession measurements.[24] For this purpose, the magnetizations of the electrodes are first aligned parallel (anti-parallel) to each other by the in-plane magnetic field $B_{\parallel}$. This is followed by a magnetic field $B_{\perp}$ perpendicular to the graphene plane forcing the spins to precess about the latter (Figure 2-c). As expected, this also yields $\Delta R \approx 4$ Ω. With $L \approx 1.15$ µm being the separation between the electrodes (center-to-center distance) and $\omega_L$ the Larmor frequency we fit our data by:

$$R_{nl} \propto \int_0^{\infty} \frac{1}{\sqrt{4\pi D_s t}} \exp(\frac{-L^2}{4D_s t}) \exp(\frac{-t}{\tau_s}) \cos(\omega_L t) \mathrm{d}t \qquad (1)$$

This gives a transverse spin relaxation time of $\tau_s \approx 180$ ps, a spin diffusion constant of $D_s \approx 0.007$ m$^2$/s and hence, a spin relaxation length ($\lambda_s = \sqrt{D_s \tau_s}$) of $\lambda_s \approx 1.1$ µm. A clear spin valve signal is also observed for the CVD BLG samples (Fig. 2-g). The origin of the signal is again confirmed by the Hanle measurements (Fig. 2-h). Using the same fitting procedure as for the SLG measurements we obtain for the BLG a spin relaxation time $\tau_s \approx 285$ ps, a spin diffusion constant of $D_s \approx 0.0063$ m$^2$/s and hence, a spin relaxation length of $\lambda_s \approx 1.35$ µm.

Next, we determine the dominant spin scattering mechanisms in CVD SLG and BLG by evaluating the functional dependence of $\tau_s$ on $\tau_p$. For the Elliott- Yafet (EY) mechanism, spin dephasing occurs during momentum scattering. Therefore, the spin relaxation time is directly proportional to the momentum scattering time ($\tau_s \propto \tau_p$).[25] On the other hand, the D'yakonov-Perel' (DP) mechanism refers to the case where spin dephasing takes place between momentum scattering events, which may result from random Bychkov-Rashba like spin-orbit fields.[26] This leads to a spin relaxation time, which is inversely proportional to the momentum scattering time ($\tau_s \propto \tau_p^{-1}$).[27] Away from the CNP, the electric field effect in graphene provides a convenient tool to correlate $\tau_s$ and $\tau_p$.[9],[28] Provided that *both* quantities show a discernible charge density dependence, such a correlation can be used to identify the limiting spin dephasing mechanism as has been demonstrated for exfoliated graphene samples. Using this approach at RT, the dominant spin scattering mechanism for exfoliated SLG spin valves with spin injection through leaky contacts has been identified to be of EY type.[7],[9] In exfoliated BLG, the DP type mechanism is dominant.[11]

We start our discussion of the Cu-CVD samples with the SLG case and note that $\tau_s$ increases with doping by ~ 35% from 175 ps to 230 ps in typical gate bias ranges. The $n$ dependence of $\tau_p$ within the Boltzman transport theory framework [28] is extracted from $\tau_p(n) = \dfrac{h\sigma}{e^2 v_F \sqrt{(n g_s g_v \pi)}}$ (Fig. 2-d), where $g_v$ and $g_s$ are the twofold valley and spin degeneracies respectively, $h$ is the Planck constant, $e$ is the electron charge and $v_F$ is the Fermi velocity. Combining both results we obtain an approximately linear scaling of $\tau_s$ with $\tau_p$, i.e. ($\tau_s \propto \tau_p$) (Fig. 2-e). In the case of CVD BLG, $\tau_s$ increases with increasing $n$ from 265 ps to 335 ps. However, $\tau_p$ shows a *decreasing* trend with increasing $n$ as extracted from $\tau_p(n) = \frac{m^*\sigma}{e^2 n}$, where $m^*$ is the effective mass of the charge carriers (Fig. 2-i).[29] Therefore, correlating $\tau_s$ and $\tau_p$, we obtain for BLG an inverse scaling (Fig. 2-j), i.e. ($\tau_s \propto \tau_p^{-1}$). These results summarize the key findings of our experiments: At RT, the typical spin parameters in CVD graphene differ neither quantitatively nor qualitatively from exfoliated graphene: $\tau_s$, $D_s$ and $\lambda_s$ are of the same order of magnitude in both systems.[6]-[12] Equally important, their charge density dependence qualitatively remains the same as exfoliated samples.[7],[9],[11] Hence, the limiting spin dephasing mechanisms at RT remain EY type and DP type in CVD SLG and CVD BLG, respectively.

These results are at first rather surprising, since CVD graphene has additional solvent residues,[30] structural differences,[18] in particular grain boundaries[17] and nanoripples[19] when compared to exfoliated graphene. Also, Cu-CVD growth typically requires high temperatures of 1000 – 1050°C. This leads to single-crystal terraces and step edges in Cu, which in turn gives rise to additional nano-ripples in graphene after transfer (inset Fig. 1-b). They are best seen in high resolution contact mode AFM images after transfer onto Si/SiO$_2$ substrates (Fig. 1-d). Such double peak structures of 0.2 – 2 nm height, ~ 100 nm width and ~ 300 nm separation are quasi-periodic across an area ($\gtrsim$ 10 µm$^2$) much larger than the actual spin valve size.[19] Assuming for example a channel area of ~ 1 × 1 µm$^2$, there will be approximately three such features present independent of their relative orientation with respect to the ferromagnetic electrodes. Thus, the growth and transfer processes cause local curvature in graphene which may affect spin-orbit coupling. In carbon nanotubes (CNT), local curvature has been shown to strongly enhance spin-orbit coupling.[2],[31] However, it is important to note that the radius of curvatures in CNT and our samples differ greatly. The average radius of curvature in quasi-periodic nano-ripples is ~ 200 nm, which leads to a much weaker spin-orbit coupling strength of ~ 3 µeV (Supplementary Information). A comparison with the intrinsic spin-orbit coupling of graphene (~ 24 µeV)[32] suggests that the nano-ripples in the present samples cannot set a limit for spin transport. The high temperature growth of graphene on the Cu surface does however have one advantage. The rather weak interaction with the underlying Cu substrates allows graphene to grow continuously crossing atomically flat terraces, step edges, and vertices without introducing defects.[33] Thus, by controlling pre-growth annealing[16] and fine tuning growth parameters,[34] it is now possible to synthesize Cu-CVD graphene with sub-millimeter grain size. The grain size of our Cu-CVD graphene is ~ 50 – 100 μm, as determined by SEM of sub-monolayer coverage graphene on the Cu foil (Fig. 1-e). This makes spin transport across grain boundaries in sub-micron size spin valves highly unlikely. Thus, under current growth conditions, neither grain boundaries nor nano-ripples, which are the two key differences of Cu-CVD graphene with respect to exfoliated graphene, have a limiting effect on spin transport. The main spin scattering mechanism in Cu-CVD samples seems to originate from the same source as in the case of spin valves based on exfoliated samples: adatoms,[35] scattering from the tunneling barrier interface[36] and the supporting substrate.[26]

Last but not least, we present spin transport measurements as a function of temperature from RT down to 5 K (Fig. 3). The temperature dependence of $\tau_s$, $\lambda_s$ and $D_s$ has been measured for three distinct doping levels: 1) at the CNP, 2) at $n \approx 7.5 \times 10^{11}/\text{cm}^2$ and 3) at $n \approx 1.5 \times 10^{12}/\text{cm}^2$. We focus our discussion on the quantity $\tau_s$. In CVD SLG spin valves, similar to results in exfoliated SLG devices,[11] we observe at all doping levels only a weak temperature dependence (Fig. 3-a). The CVD BLG, on the other hand, shows a more complex temperature dependence of $\tau_s$, which differs strongly between the CNP and the high carrier densities (Fig. 3-b). At high doping, $\tau_s$ in CVD BLG is only weakly temperature dependent. However, at the CNP $\tau_s$ shows a sharp increase from 260 ps to 360 ps between 250 K and 200 K. In contrast, for temperatures above 250 K and below 200 K, $\tau_s$ varies again only weakly with temperature at the CNP. This non-monotonic temperature dependence at the CNP is typical also for our exfoliated bilayer devices.[11] Finally, we discuss the $n$ dependence of $\tau_s$ and $\lambda_s$ in CVD samples at 5 K. Similar to the RT case, at low temperature (LT) $\tau_s$ increases with increasing $n$ in SLG (Fig. 4-a). This implies that in SLG the main scattering mechanism remains of the EY type even at LT. However, in CVD BLG, $\tau_s$ decreases with increasing $n$ in contrast to RT (Fig. 4-b). This behavior becomes noticeable for temperatures below 200 K, but is most pronounced at the lowest measured temperature ($T = 5K$). While this qualitative change of the charge density dependence of $\tau_s$ at low temperature is not yet understood, it is bilayer specific. A very similar behavior has also been observed previously in exfoliated BLG samples.[11],[12] Thus, comparing our CVD graphene results with results obtained in exfoliated graphene spin valves, we conclude that the temperature and the carrier density dependence of $\tau_s$ is comparable in both systems.

In conclusion, we have demonstrated spin injection, spin transport and spin detection in Cu-CVD SLG and BLG samples. The key spin transport parameters such as $\tau_s$ and $\lambda_s$ have been measured as a function of charge carrier density and temperature. They are comparable to what has been already reported in both exfoliated SLG and exfoliated BLG samples making Cu-CVD graphene a promising candidate for possible large scale spintronic applications. We have also discussed the importance of Cu-CVD graphene specific quasi-periodic arrays of nanoripples in spin transport. While in current samples the local curvature is too small to enhance the spin-orbit coupling significantly, such quasi-periodic nanoripple arrays may provide intriguing opportunities in controlling spin currents through spin-orbit coupling due to local curvature and local strain.

**Acknowledgement.** This work is supported by the Singapore National Research Foundation grants NRF-RF2008-07, and NRF-CRP (R-143-000-360-281), NUS/SMF horizon grant, US Office of Naval Research (ONR and ONR Global), by NUS NanoCore, by DFG through FOR 912 and by JARA-FIT.

**Supporting Information Available:** This material is available free of charge via the Internet at http://pubs.acs.org.

FIGURE CAPTIONS

Figure 1 (a) Optical image of CVD single and bilayer graphene on Si/SiO$_2$ substrate (300 nm SiO$_2$ thickness) with their respective Raman spectra. Black and red circles indicate the Raman spectroscopy locations. Blue arrows point to low density wrinkles typical for CVD graphene films. (b) Schematics for a graphene based non-local spin-valve together with a possible configuration of quasi-periodic nano-ripples in a spin-valve. (c) Bottom: Optical image of a 5 × 5 device array. CVD graphene allows the fabrication of large arrays of identical lateral spin valves. Top: Scanning electron micrograph of CVD SLG spin sample with multiple non-local spin valve devices. Electrode widths range from 0.3 µm to 1.2 µm. (d) High resolution contact mode AFM image of CVD graphene after transfer onto Si/SiO$_2$ wafer revealing the presence localized nanoscale ripples of high density. (e) SEM image of sub-monolayer graphene coverage on Cu.

Figure 2 (a) Conductivity of CVD SLG at RT and at $T = 5$ K as a function of carrier density with a strong asymmetry between electron and hole doped region. (b) Bi-polar spin signal obtained near the charge neutrality point. (c) Hanle spin precession measurement confirms the spin signal in b). (d) Both $\tau_s$ and $\tau_p$ increase with increasing electron carrier density. (e) Linear dependence of $\tau_s$ and $\tau_p$ showing that EY like spin scattering is dominant in CVD SLG. (f) Conductivity of CVD BLG at RT and at $T = 5$ K as a function of carrier density. (g & h) Spin valve and spin precession measurements in CVD BLG, respectively. (i) Electron carrier density dependence of $\tau_s$ and $\tau_p$ at RT. (j) Scaling of $\tau_s$ with $\tau_p$ indicates DP type spin scattering as the dominant spin scattering mechanism in CVD BLG.

Figure 3 (a) $T$ dependent $\tau_s$ and $\lambda_s$ are shown for CVD SLG for three different electron carrier densities. (b) The $T$ dependences of $\tau_s$ have different behavior for different doping levels in CVD BLG. $\lambda_s$ depends very weakly on $T$, but its $n$ dependence is much weaker than for CVD SLG. $\lambda_s$ is observed to be very weakly dependent on temperature for fixed carrier densities in both CVD SLG and BLG, since different temperature dependence trends of $\tau_s$ and $D_s$ almost suppress each other in both systems (see Supplementary).

Figure 4 (a) Charge carrier density dependence of $\tau_s$ and $\lambda_s$ at RT and at 5 K for CVD SLG. (b) Charge carrier density dependence of $\tau_s$ and $\lambda_s$ at RT and 5 K for CVD BLG. Note that the carrier density dependence of $\tau_s$ of CVD BLG at 5 K shows an opposite trend compared to RT.

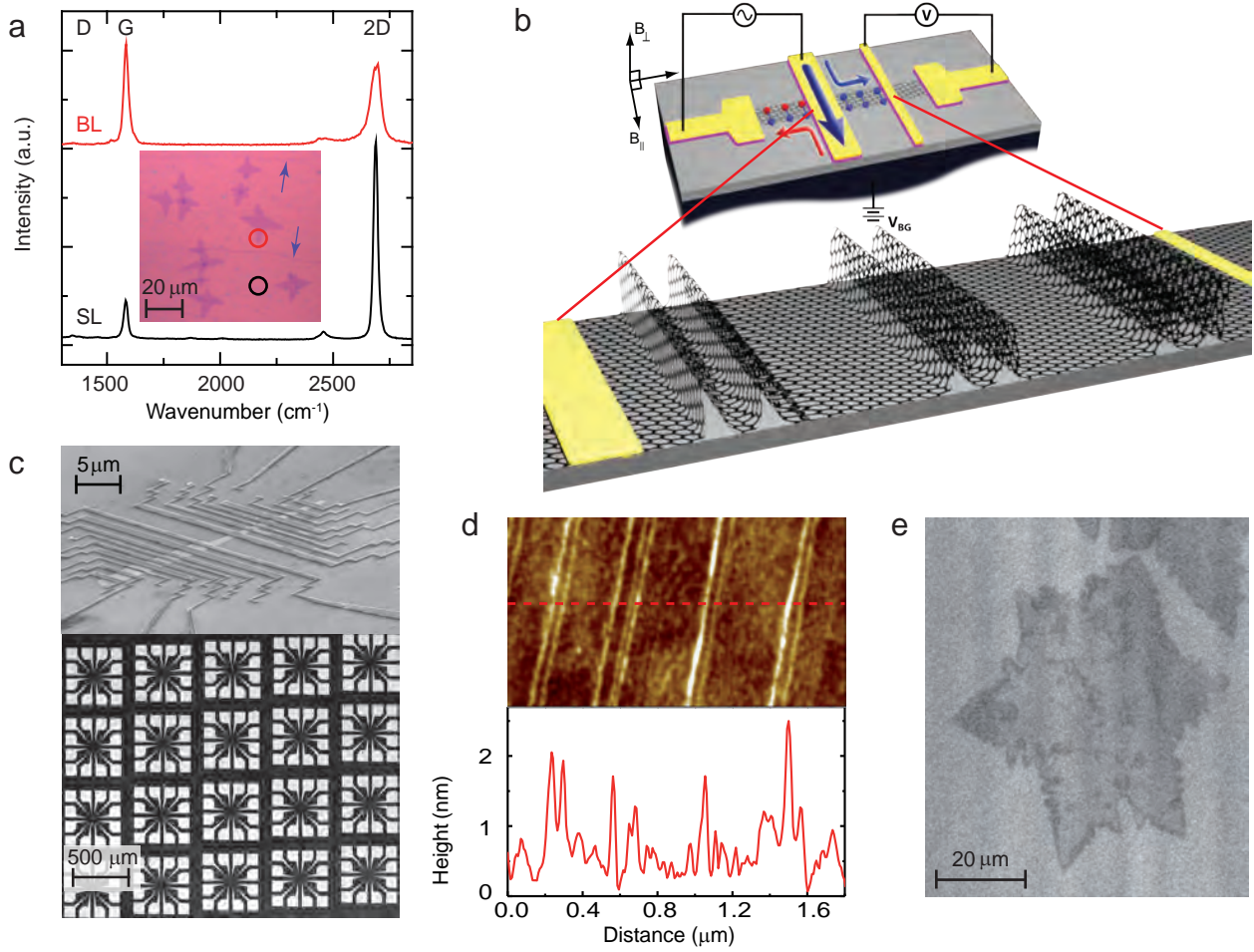

Figure 1: A. Avsar *et al.*,

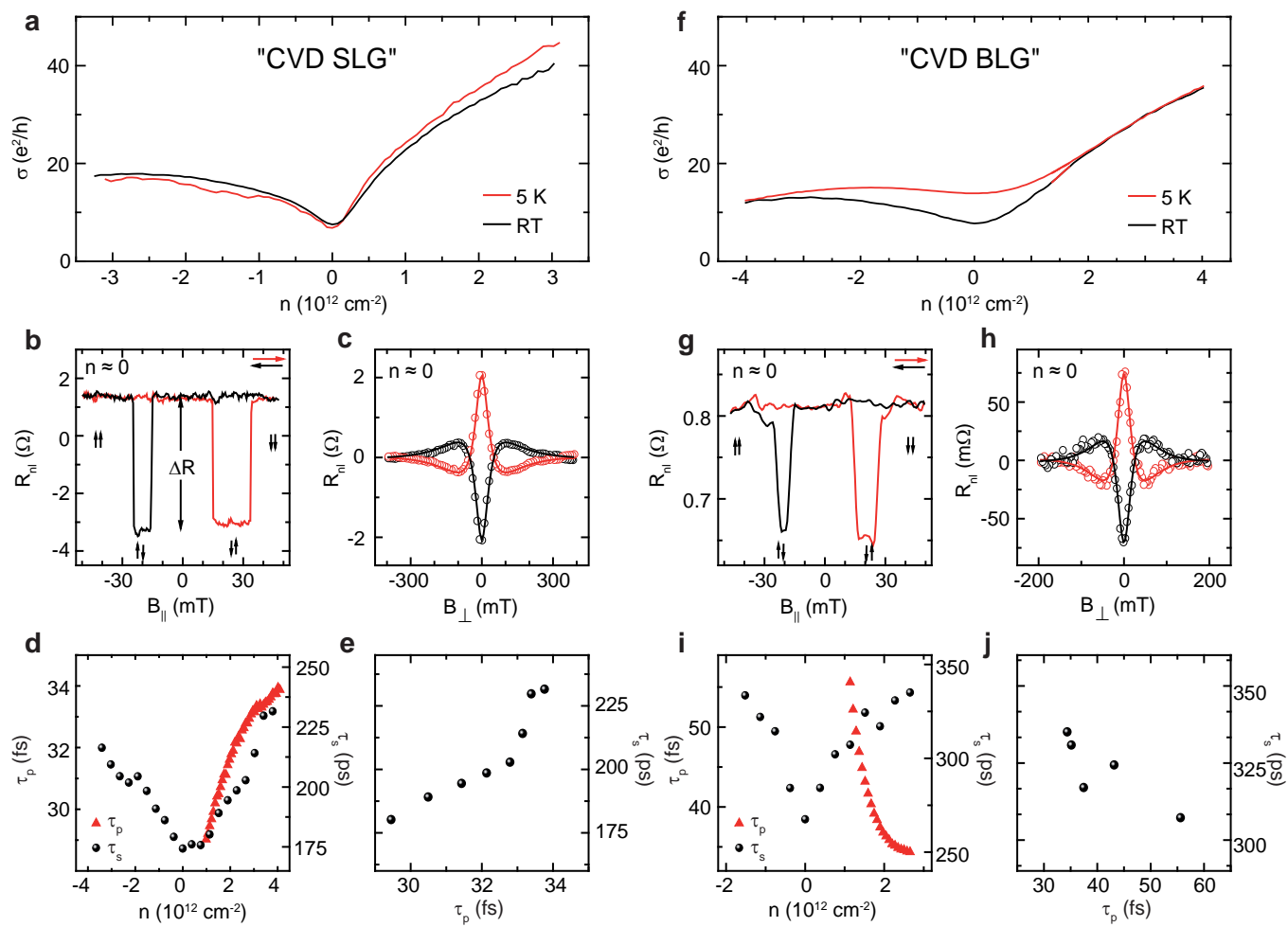

Figure 2: A. Avsar *et al.*,

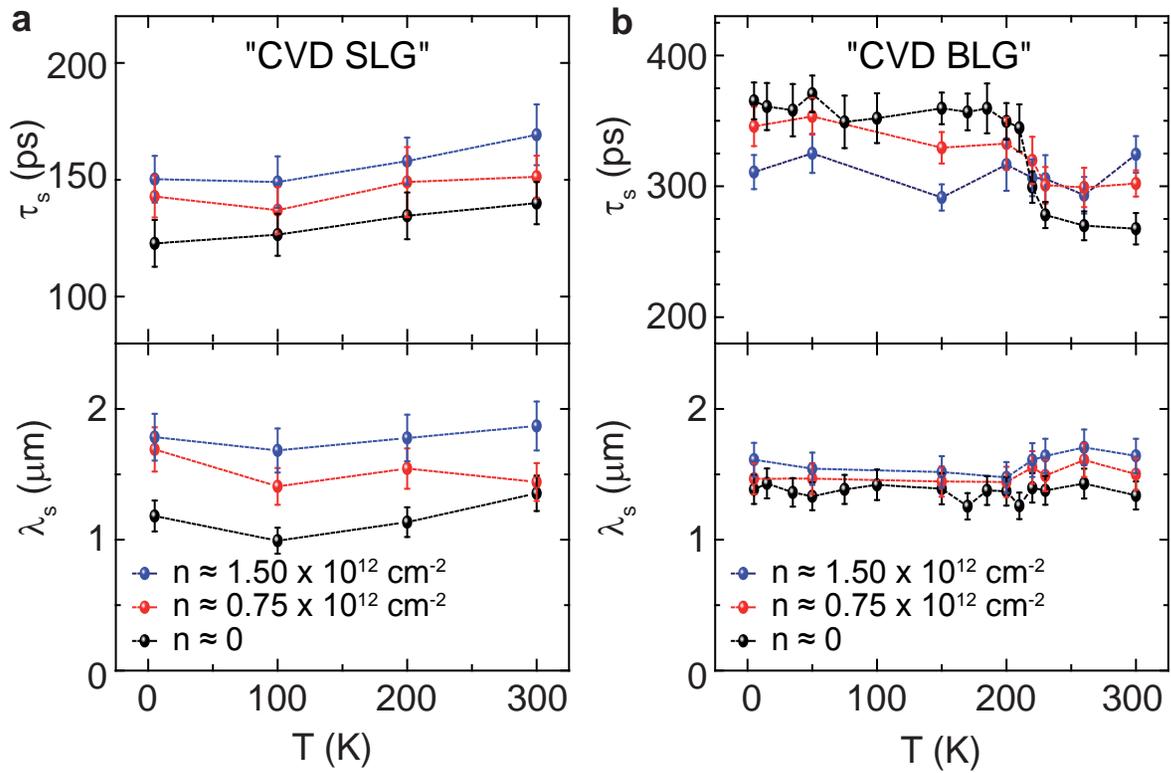

Figure 3: A. Avsar *et al.*,

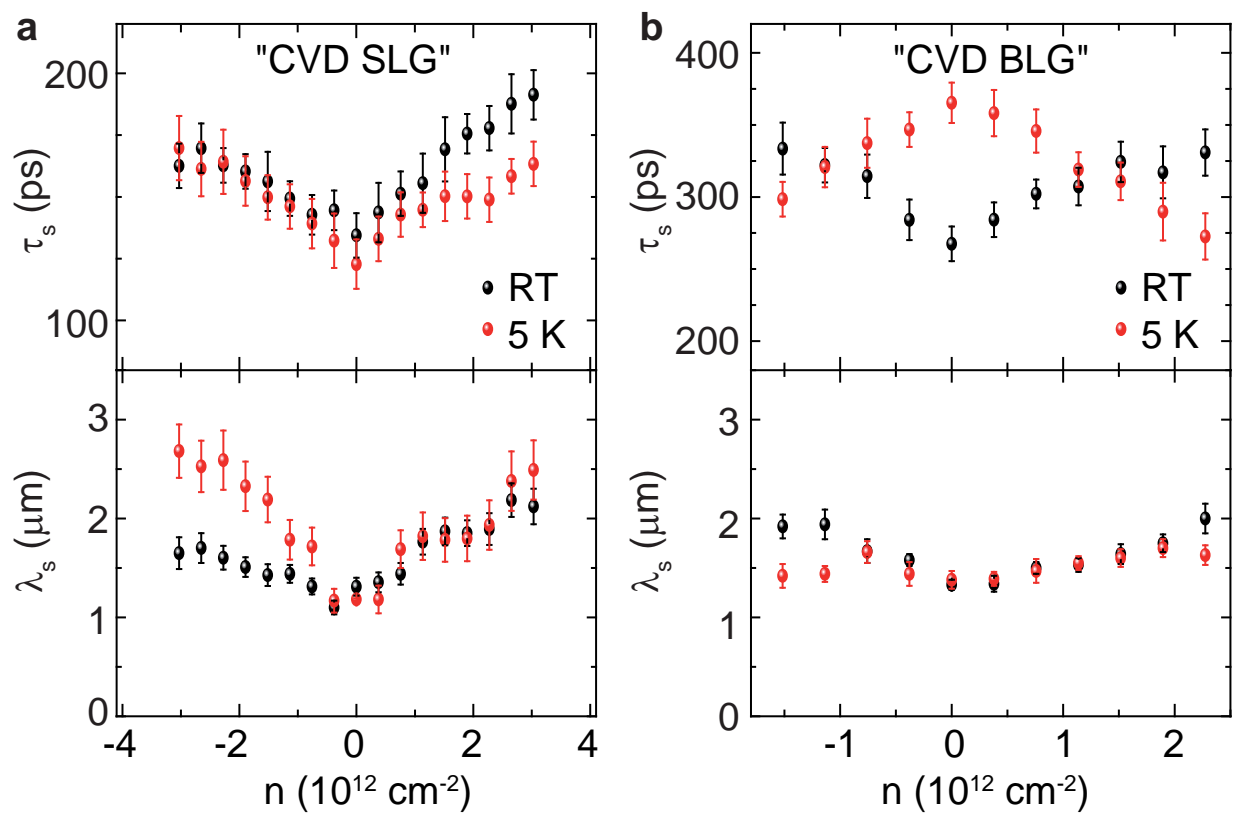

Figure 4: A. Avsar *et al.*,

# Towards wafer scale fabrication of graphene based spin valve devices

*Ahmet Avsar[1,]\*, Tsung-Yeh Yang[2, 3,]\*, Su-Kang Bae[4,]\*, Jayakumar Balakrishnan[1], Frank Volmer[2,3], Manu Jaiswal[1], Zheng Yi[1, 4], Syed Rizwan Ali[2,3], Gernot Güntherodt[2,3], Byung-Hee Hong[4,†], Bernd Beschoten[2,3,†], and Barbaros Özyilmaz[1,5,6,†]*

[1]Graphene Research Center & Department of Physics, 2 Science Drive 3, National University of Singapore, Singapore 17542, Singapore

[2]II. Institute of Physics, RWTH Aachen University, 52074 Aachen, Germany

[3]JARA: Fundamentals of Future Information Technology, 52074 Aachen, Germany

[4]SKKU Advanced Institute of Nanotechnology (SAINT) and Center for Human Interface Nanotechnology (HINT), Sungkyunkwan University. Suwon, 440-746, Korea

[5]Nanocore, 4 Engineering Drive 3, National University of Singapore, Singapore 117576, Singapore

[6]NUS Graduate School for Integrative Sciences and Engineering (NGS), Singapore.

\*These authors contributed equally to this work

[†]Corresponding authors: barbaros@nus.edu.sg, bernd.beschoten@physik.rwth-aachen.de, byunghee@skku.edu

# 1. Carrier density (n) dependence of spin signal (ΔR), spin diffusion constant ($D_s$) and spin injection efficiency (P) in CVD single layer graphene (SLG) and bilayer graphene (BLG).

The $n$ dependence of $\Delta R$, $D_s$, and $P$ is studied in CVD SLG and CVD BLG samples at room temperature (RT) (Fig. S1). In CVD SLG, $\Delta R$ exhibits only a weak $n$ dependence with a minimum near the charge neutrality (CNP) and saturation away from CNP. This behavior is indicative of pin-holes in the MgO barrier.[1] We note that even though we do not use a $TiO_2$[1] as buffer layer, we do still obtain a uniform, continuous MgO layer. The $D_s$ shows strong $n$ dependence and increases with increasing $n$ by more than 300% in typical gate bias ranges.

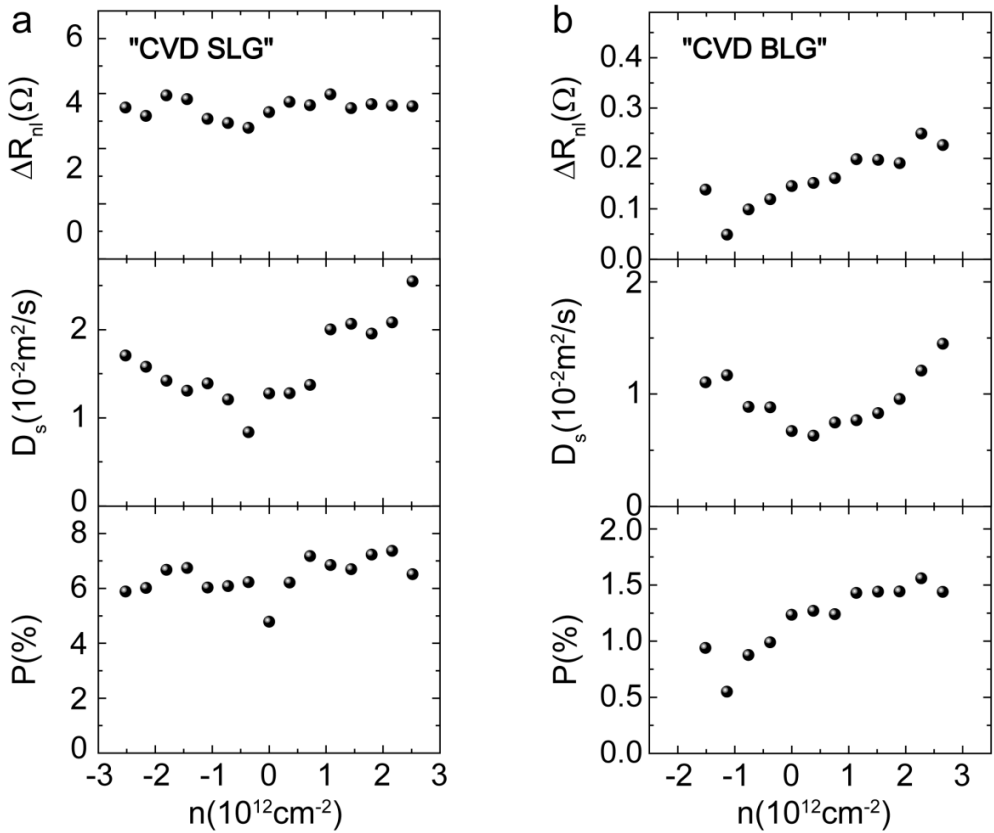

Figure S1. The $n$ dependence of $\Delta R$, $D_s$, and $P$ in (a) CVD SLG and (b) CVD BLG at RT.

The $n$ dependence of $P$ is extracted from $\Delta R = \dfrac{P^2 \lambda_S e^{(-L/\lambda_S)}}{w\sigma}$, where $\lambda_s$ is the spin relaxation length, $w$ is the width of graphene strip and $\sigma$ is the conductivity of graphene sheet. Similar to $\Delta R$, $P$ shows a weak $n$ dependence. In the CVD BLG, we observe a strong *n* dependence of the $D_s$ as in CVD SLG. The $\Delta R$ and the *P* in CVD BLG show a distinct behavior on the hole side, but such differences are sample dependent. We attribute the relatively smaller $\Delta R$ and *P* in CVD BLG compared to CVD SLG to the quality of the barrier instead of the number of graphene layers.

## 2. Carrier density (n) dependence of conductivity (σ), spin relaxation time ($\tau_s$) and spin relaxation length ($\lambda_s$) in exfoliated SLG and BLG.

Prior to the *n* dependent spin precession measurements, the conductivity σ of exfoliated SLG and BLG is characterized as a function of *n* at RT (Fig. S2). We observe a strong distortion on the hole side, similar to the recent report on graphene field effect transistor with ferromagnetic electrodes.[2] The $n$ dependence of the $\tau_s$ and $\lambda_s$ in exfoliated samples at RT is qualitatively comparable to what is observed in their CVD counter parts (Fig. 4). Note that changes of up to 50 % are observed even within nominally identical samples. The strong change of the $n$ dependence of $\tau_s$ at low temperature is observed in both exfoliated BLG and CVD BLG but not in SLG samples.[3]-[4]

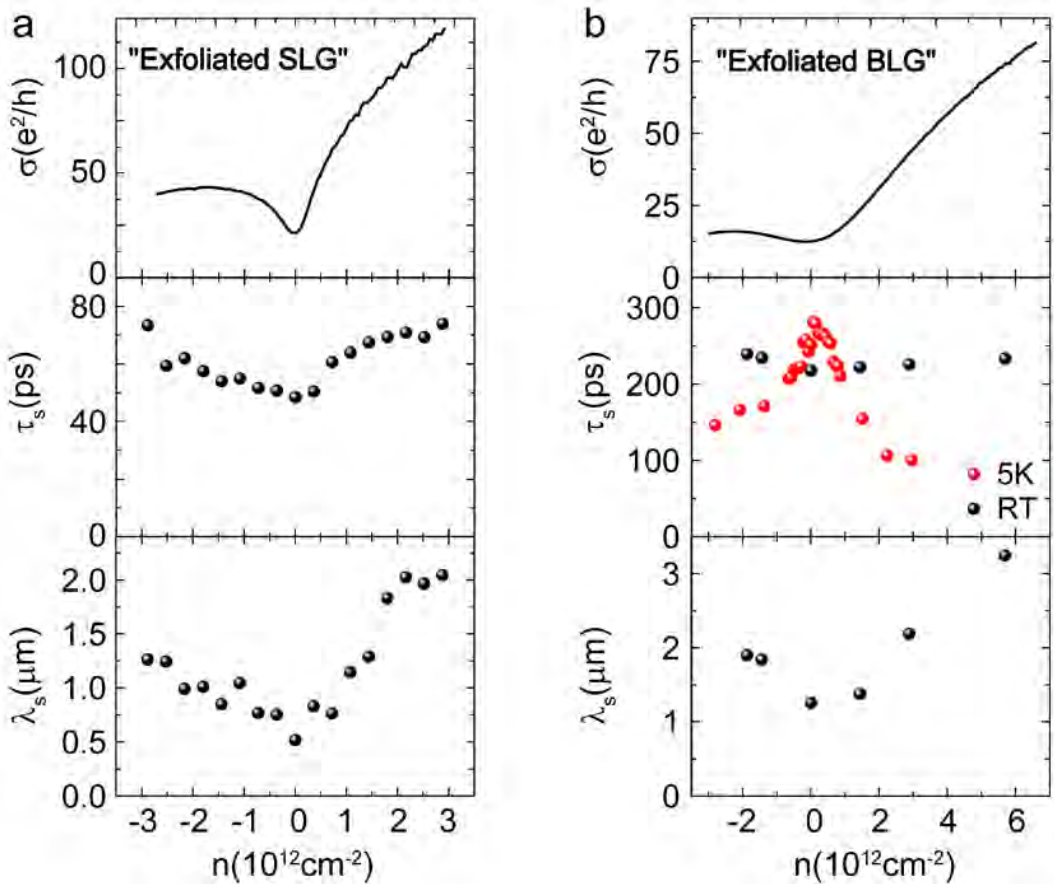

Figure S2. The *n* dependence of *σ*, $\tau_s$ and $\lambda_s$ in exfoliated (a) SLG and (b) BLG at RT. The spin transport parameters and their gate bias dependence are comparable in exfoliated graphene and CVD graphene at both RT and LT.

## 3. Estimate of the nanoripple induced spin-orbit coupling (SOC) strength

Nanoripples in CVD graphene do have an impact on the charge transport.[5] However, in spin transport the influence of such nanoripples greatly depends on the curvature in the nanoripples.[6] In CNT with a small curvature an enhancement of spin orbit of up to 0.32 meV has been reported.[7] We estimate the spin-orbit coupling strength induced by nanoripples in CVD graphene. Figure S3 shows the calculated curvature for such nanoripples with an average radius of curvature of $R \sim 200$ nm determined by a Gaussian fit of a single AFM trace across a nanoripple. The smallest curvature of $\sim 100$ nm is observed at the peak of the nanoripple. Following Ref. (6) and using their parameters, we estimate the curvature induced spin-orbit coupling strength $\Delta_{curv}$ as

$$\Delta_{curv} = \frac{\Delta(V_{pp\sigma} - V_{pp\pi})}{V_1}\left(\frac{a}{R}\right)\left(\frac{V_1}{V_2}\right)^2, \text{ where } V_1 = \frac{(E_s - E_p)}{3}, \quad V_2 = \frac{(2V_{pp\sigma} + 2\sqrt{2}V_{sp\sigma} + V_{ss\sigma})}{3}.$$

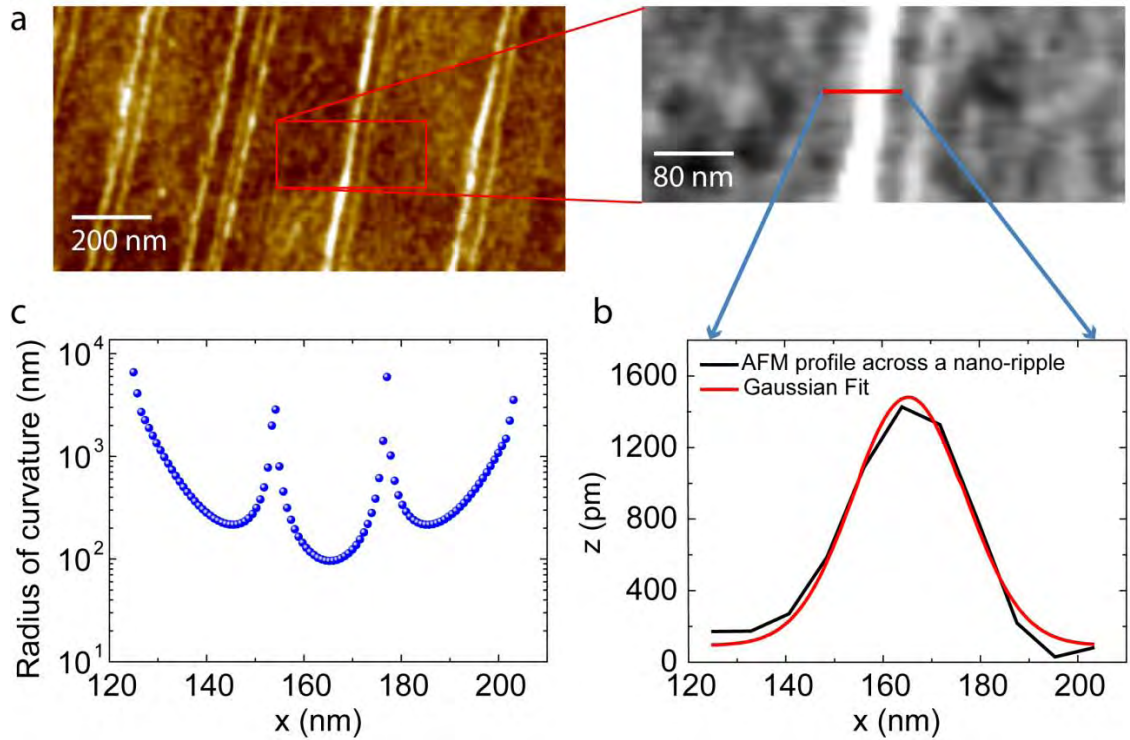

Figure S3. Estimate of the SOC strength induced by nanoripples in CVD graphene. (a) The AFM image of nanoripples in Cu-CVD graphene. (b) The Gaussian fit to the nanoripple for determining the radius of curvature $R$. (c) Radius of curvature determined from the Gaussian fit to the nanoripple.

Here, $\Delta = 12$ meV is the strength of the intra-atomic spin–orbit coupling of carbon, $V_{pp\sigma} = 5.38$ eV, $V_{pp\pi} = -2.24$ eV, $V_{sp\sigma} = 4.2$ eV, $V_{ss\sigma} = -3.63$ eV are the hoping amplitudes between 2s, $2p_x$ and $2p_y$ of the $\sigma$ band and $2p_z$ of the $\pi$ band, $(E_s - E_p) = 7.41$ eV is the energy difference between the 2p and the 2s atomic orbitals, $a = 1.42$ Å is the nearest neighbor distance, $V_1 = 2.47$ eV, $V_2 = 6.33$ eV and $R \sim 200$ nm is the radius of curvature of the nano-ripples in our samples. The estimate gives a nanoripple induced spin-orbit coupling strength of 3.3 µeV. This value is significantly less than the predicted intrinsic SOC in pristine graphene,[8] and therefore, in current samples, would not give rise to the limiting spin scattering mechanism. However, it should be noted that the size of nanoripples depends strongly on the various pre-annealing, cooling rate and growth conditions. Thus controlling these growth factors could enable the control of the radius of curvature of these nanoripples. We estimate that the radius of curvature needed to affect spin relaxation times of the order of $\leq 10$ ns has to be $\leq 30$ nm.